\documentclass[aps,prl,twocolumn,showpacs,groupedaddress,preprintnumbers]{revtex4}

\usepackage{graphicx}

\begin{document}

\preprint{MIFP-07-13}

\title{Constraint from $D$-$\bar D$ Mixing in Left-Right Symmetric Models
}


\author{Bhaskar Dutta}
\author{Yukihiro Mimura}
\affiliation{
Department of Physics, Texas A\&M University,
College Station, TX 77843-4242, USA}


\date{August 22, 2007}

\begin{abstract}
We study the constraint arising from the recently observed $D$-$\bar
D$ mixing in the context of supersymmetric models with left-right
symmetry. In these models, the supersymmetric contributions in the
mixing amplitudes of $D$-$\bar D$, $K$-$\bar K$ and $B$-$\bar B$ are
all correlated. We compare the constraint from the $D$-$\bar D$
mixing with the $K$-$\bar K$ mixing and find that the $D$-$\bar D$
mixing constrains the maximal supersymmetric contribution to the
$B_s$-$\bar B_s$ mixing amplitude.
The maximal supersymmetric contribution 
can allow a large CP phase of $B_s$-$\bar B_s$ mixing
which agrees with
the recent measurement of the CP asymmetry of $B_s \to J/\psi \phi$ decay.
%
\end{abstract}

%
\pacs{12.10.-g, 12.15.Ff}
%

\maketitle


Recently, BaBar and Belle have observed signals of $D^0$-$\bar D^0$
mixing \cite{Aubert:2007wf}. The HFAG \cite{hfag} interpretation of
the current data gives us
\begin{equation}
x_D = 8.7^{+3.0}_{-3.4} \times 10^{-3},\ \ y_D = (6.6\pm 2.1) \times 10^{-3},
\end{equation}
where $x_D=\Delta M_D/\Gamma_D$ and $y_D=\Delta \Gamma_D/(2
\Gamma_D)$, $\Gamma_D$ is the average decay width of two neutral $D$
meson mass eigenstates. The mass difference of $D^0$-$\bar D^0$ is
obtained as
\begin{equation}
\Delta M_D = (1.4\pm 0.5) \times 10^{-11}\ \mbox{MeV}.
\end{equation}
This new data can constrain new physics such as supersymmetry (SUSY)
in the similar way as the traditional constraint from
the $K$-$\bar K$ mixing data~\cite{Ciuchini:2007cw}.

In SUSY models, the flavor degeneracy is often assumed in squark and
slepton mass matrices to suppress flavor changing neutral currents
(FCNC)~\cite{Gabbiani:1988rb}. The flavor violation effects in the
sfermion mass matrices can only come from the evolution of
renormalization group equations (RGE).
If this is the case, the
flavor violation highly depends on the unification scenario of
quarks and leptons. In the minimal extension of SUSY standard model
(MSSM), the induced FCNCs from RGE effects are not large in the
quark sector, but sizable effects can be generated in the lepton
sector since the neutrino mixings are large \cite{Borzumati:1986qx}.
In quark-lepton unified models, the loop effects due to the large
neutrino mixings can induce sizable effects also in the quark
sector. Therefore, it is important to investigate the FCNC effects
to obtain a footprint of the unification models.

Left-right symmetric model construction is an interesting candidate
to unify matter (including right-handed neutrino) in the gauge group
$SU(3)_c \times SU(2)_L \times SU(2)_R \times
U(1)_{B-L}$~\cite{moha}. The left-right parity is broken
spontaneously, and the hypercharge arises from a linear combination
of $U(1)_{B-L}$ and the $U(1)$ subgroup of $SU(2)_R$. This
left-right symmetric branch can be easily unified in $SO(10)$ grand
unified models.
In the SUSY version of left-right symmetric models, the box diagrams
for meson mixing ($K$-$\bar K$, $B$-$\bar B$ and $D$-$\bar D$) can
be enhanced by gluino contribution. Therefore, the newly observed
$D$-$\bar D$ mixing can be an important probe for left-right
symmetric models.
Further, in such models,   $D$-$\bar D$, $K$-$\bar K$, $B_d$-$\bar
B_d$ and $B_s$-$\bar B_s$ mixing amplitudes get correlated. This
creates an interesting opportunity for cross-checking these models,
since the most interesting observation from the sizable SUSY
contribution will be the phase of $B_s$-$\bar B_s$ mixing, which can
be measured by $B_s \to J/\psi \phi$ decay.

The mass difference of  $B_s$-$\bar B_s$ (the absolute value of the
mixing amplitude $M_{12}$) has been measured
\cite{Abulencia:2006ze}, and the measurement is consistent with the
Standard Model (SM) prediction.
Therefore, if there is a sizable SUSY contribution, the phase of
$B_s$-$\bar B_s$ mixing (argument of the amplitude) must be large.
The CP asymmetry of the $B_s$ decay is being measured and the
current result is \cite{Abazov:2007zj}
\begin{equation}
2 \beta_s = -0.70^{+0.47}_{-0.39} \ ({\rm rad}),
\label{beta_s}
\end{equation}
while the SM prediction is $\sim 0.03-0.04$. If this
result holds in future, then it will indicate an existence of new
physics.

The SUSY contribution to the $B_s$-$\bar B_s$ mixing is related to
the 23 off-diagonal elements of the squark mass matrices, which may
be large since it can be related to the large atmospheric mixing.
On the other hand,
it is hard to predict the amount of the SUSY contribution to the
$K$-$\bar K$ and $D$-$\bar D$ mixings due to cancellation.
However,
we can show that cancellations for both $K$-$\bar K$ and $D$-$\bar
D$ mixings are not allowed simultaneously when the non-universal
terms in squark mass matrices originate from left-right symmetric
models. Consequently, the recent observation of the $D$-$\bar D$
mass difference restricts the amount of SUSY contribution, and thus,
it also restricts the phase of $B_s$-$\bar B_s$ mixing.
In this Letter, we will show how to obtain the constraint of SUSY
contribution from the $D$-$\bar D$ mixing, and study the correlation
of the constrained phase of $B_s$-$\bar B_s$ mixing to other
measurements, e.g., phase of $B_d$-$\bar B_d$ mixing, 
in left-right symmetric models.

In left-right symmetric models, the Lagrangian is invariant under
the exchange $Q_L \leftrightarrow Q_R^c$, where $Q_L$ is $SU(2)_L$
doublet and $Q_R^c$ is $SU(2)_R$ doublet which contains conjugate of
right-handed up- and down-type quarks $U^c$, $D^c$. As a result,
Yukawa couplings are given by symmetric matrices. The squark
matrices are given at unification scale as \cite{Dutta:2006zt}
\begin{equation}
M_F^2 = m_0^2 \left( {\bf 1} - \kappa \, U_F
\left(\begin{array}{ccc}
k_1 && \\
& k_2 & \\
& & 1
\end{array}\right)
U_F^\dagger \right) ,
\end{equation}
where $U_F$ is a unitary matrix and $F$ denotes $Q$, $U^c$ and $D^c$.
In the original basis where we respect the left-right symmetry,
$U_F$ is common for $Q$, $U^c$ and $D^c$.
We note that the squark mass matrices are given in the notation:
$(M_{Q}^2)_{ij} \tilde Q_i \tilde Q^\dagger_j
+(M_{U}^2)_{ij} \tilde U_i^c \tilde U^{c\dagger}_j
+(M_{D}^2)_{ij} \tilde D_i^c \tilde D^{c\dagger}_j
$.
The non-universal part of squark mass matrices
can be generically parameterized by the $\kappa$ term.
We consider that the $\kappa$ term is generated from a loop diagram
in the form $\propto f f^\dagger$. Here, $f$ is
the quark Majorana coupling $f (Q_L Q_L \Delta_{qq} + Q^c_R Q^c_R \Delta_{qq}^c)$,
which can be unified into the neutrino Majorana coupling
in a $SO(10)$ model.
In general,
$U_F$ is parameterized as $U_F = P U_q$
where $P$ is a diagonal phase matrix
 and $U_q$
includes 3 mixing angles ($\theta_{ij}^q$) and 1 phase ($\delta^q$).
We parameterize $U_q$ in the basis where the down-type quark Yukawa
matrix is diagonal.
The mixing angles and a phase are parameterized in the same
convention as the CKM matrix. If we consider the type II seesaw
scenario \cite{Schechter:1980gr} in a $SO(10)$ model, $k_2$
corresponds to the ratio of neutrino mass squared and $U_F$ is the
neutrino mixing matrix in the basis where the charged-lepton mass
matrix is diagonal, and thus $\theta_{ij}^q$ correspond to neutrino
mixing angles when both charged-lepton and down-type quark mass
matrices are simultaneously diagonalized. In general,
$\theta_{ij}^q$ are not necessarily exactly same as the neutrino
mixings. The Yukawa matrices for up- and down-type quarks ($Y_u$ and
$Y_d$) are given as
\begin{equation}
Y_u = V_{\rm CKM}^{\rm T} Y_u^{\rm diag} P_u V_{\rm CKM}, \quad
Y_d = Y_d^{\rm diag} P_d,
\end{equation}
where $P_{u,d}$ are diagonal phase matrices.

We can calculate the off-diagonal elements
of the squark mass matrices $\delta_{ij} \equiv (M_F^2)_{ij}/m_0^2$
in the above notation.
\begin{eqnarray}
\!\!|\delta_{12}^d | &\!\!\simeq&\!\! \kappa
 \left| \frac12 k_2 \sin2\theta_{12}^q \cos \theta_{23}^q +
    e^{i\delta^q} \sin\theta_{13}^q \sin \theta_{23}^q \right|,
\label{delta12} \\
\!\!|\delta_{13}^d | &\!\!\simeq&\!\! \kappa
 \left| \frac12 k_2 \sin2\theta_{12}^q \sin \theta_{23}^q -
    e^{i\delta^q} \sin\theta_{13}^q \cos \theta_{23}^q \right|,
\label{delta13} \\
\!\!|\delta_{23}^d | &\!\!\simeq&\!\! \frac12 \kappa \sin2\theta_{23}^q \,,
\label{delta23}
\end{eqnarray}
where superscript $d$ stands for that it is given
in the basis where the down-type Yukawa matrix is diagonal.
These quantities enter into the  calculation of
$K$-$\bar K$, $B_d$-$\bar B_d$ and $B_s$-$\bar B_s$
mixing amplitudes.

When a flavor degeneracy is assumed at the unification scale and
only the MSSM RGE is considered, the chargino diagram contribution
dominates the SUSY contribution. However, if the flavor violation is
induced by a loop diagram at the unification scale (as discussed before), the gluino diagram
can generate the dominant contribution. This contribution to the
mixing amplitude $M_{12}^{\tilde g}$ can be written in the following
mass insertion form
\begin{equation}
\frac{M_{12}^{\tilde g}}{M_{12}^{\rm SM}}
\simeq
a\, [(\delta_{LL}^{\tilde d})_{ji}^2+ (\delta_{RR}^{\tilde d})_{ji}^2]
- b \, (\delta_{LL}^{\tilde d})_{ji} (\delta_{RR}^{\tilde d})_{ji},
\label{gluino-contribution}
\end{equation}
($ji = 21, 31, 32$ for $K$-$\bar K$, $B_d$-$\bar B_d$ and $B_s$-$\bar B_s$,
respectively)
where  $a$ and $b$ depend on squark and gluino masses, and
$\delta_{LL,RR}^{\tilde d} = (M^2_{\tilde d})_{LL,RR}/\tilde m^2$ ($\tilde m$
is an averaged squark mass). The matrix $M^2_{\tilde d}$ is a
down-type squark mass matrix $ (\tilde Q, \tilde D^{c\dagger})
M^2_{\tilde d} (\tilde Q^\dagger, \tilde D^c)^{\rm T} $ in the basis
where the down-type quark mass matrix is real (positive) diagonal.
When squark and gluino masses are less than 1 TeV, $a \sim O(1)$ and
$b \sim O(100)$. We also have contributions from $\delta_{LR}^d$,
but we neglect them  since they are suppressed by $(m_b/m_{\rm
SUSY})^2$.
It is worth noting that the left-right symmetric boundary conditions
give much larger SUSY contribution since both off-diagonal elements
for $LL$ and $RR$ are large and $b \gg a$ in the mass insertion
formula.

When $LL$-$RR$ contributions are dominant, the phases in $P$ are
cancelled due to $(M^2_{\tilde d})_{LL} = M_F^2$ and $(M^2_{\tilde
d})_{RR} = (M_F^2)^{\rm T}$ (when we neglect the RGE effects). The
phase of the mixing amplitude is generated from the phases in $P_d$.
Since there is no constraint for the phases in $P_d$, the phase of
the mixing amplitude is free. However, there are only two physical
phases in $P_d$ and therefore the phases of the SUSY contributions for
$K$-$\bar K$, $B_d$-$\bar B_d$ and $B_s$-$\bar B_s$ are correlated.
We will show the impact of this correlation  later.

The gluino contribution for the $D$-$\bar D$ mixing is  obtained
when we change $\tilde d$ to $\tilde u$, but it needs to be written
in the basis where the up-type quark Yukawa matrix is diagonal.
The important quantity for the $D$-$\bar D$ mixing is
$\delta_{12}^u$ (in $Y_u$ diagonal basis), which can be written as
\begin{eqnarray}
[V_{\rm CKM}^* (\delta^d) V_{\rm CKM}^{\rm T}]_{12} \sim \delta_{12}^d + V_{us} \delta_{22}^d,
\end{eqnarray}
up to the $P_u$ phase
 ($P_u$ phase gives just an overall phase of $\delta_{12}^u$
and it is not important for the cancellation
since the short-distance SM contribution of $D$-$\bar D$ is small.),
and $\delta_{22}^d \simeq \kappa \sin^2 \theta_{23}^q$.
Therefore, when $\kappa \sin^2 \theta_{23}^q$ is large,
both $K$-$\bar K$ ($\delta_{12}^d$) and $D$-$\bar D$ ($\delta_{12}^u$)
SUSY contribution
cannot be cancelled away
simultaneously.

In Fig.1, we show the maximal value for $\kappa$ allowed by the
experimental results  for $K$-$\bar K$ and $D$-$\bar D$ mixings as a
function of $\sin \theta_{13}^q$. We use $\sin^2 \theta_{23}^q =
1/2$, $\tan^2 \theta_{12}^q = 0.4$, $k_1 = 0$ and $k_2 = 0.05$. The
SUSY parameters are chosen to be $m_0 = 1$ TeV, $m_{1/2} = 300$ GeV
(gaugino mass), $A_0 = 0$ (trilinear scalar coupling) and $\tan
\beta_H = 10$ (ratio of Higgs vacuum expectation values).
The phase $\delta^q$ and the other phases are chosen to make the
$\kappa$ value maximal.
In the usual convention, $\sin \theta_{13}^q$ is positive since its
negative value can be redefined by rephasing $\delta^q$. But, in
order to show the figure simply, we also use negative $\sin
\theta_{13}^q$ as a convention.
%
%
The $K$-$\bar K$ ($\delta_{12}^d$) is cancelled
at $\sin \theta_{13}^q \sim - \frac12 k_2 \sin2\theta_{12}^q \cot\theta_{23}^q$
and $D$-$\bar D$ ($\delta_{12}^u$) is cancelled
at
$\sin \theta_{13}^q \sim \pm \sin \theta_{23}^q V_{us}$.
Due to the fact that phases in $P$ are free, $D$-$\bar D$ ($\delta_{12}^u$)
can be cancelled for both positive and negative $\theta_{13}$.

For most of Fig.1, the $D$-$\bar D$
constraint, using the recent experimental result, is weaker than the $K$-$\bar K$ constraint.
However,
the $D$-$\bar D$ mixing is important at the $K$-$\bar K$ cancellation region
($\delta_{12}^d \to 0$).
%
%
As a result,
the newly observed $D$-$\bar D$ mixing can restrict the
maximal SUSY contribution to the $B$-$\bar B$ mixing.
%
{}From Fig.1, we see that the maximal SUSY contribution is obtained
at the $K$-$\bar K$ cancellation region after satisfying the $D$-$\bar D$ constraint.

\begin{figure}[t]
 \center
 \includegraphics[viewport = 25 24 280 220,width=8.0cm]{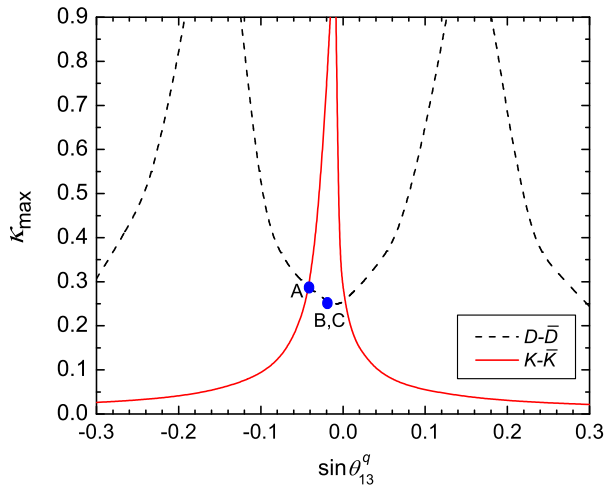}
 \caption{
Maximal values for $\kappa$ are shown as a function of the angle
$\theta_{13}^q$ from the constraints of $K$-$\bar K$ and $D$-$\bar
D$ mixing separately. Cancellation happens for the SUSY contribution
when the maximal $\kappa$ is large. The points A, B, C correspond to
the solutions illustrated in Fig.2. } \vspace{-0.5cm}
\end{figure}

We can classify the solution for fitting the of $K$-$\bar K$ mixing
amplitude in the following three cases, which are illustrated in
Fig.2. The $K$-$\bar K$ mixing amplitude is given as $M_{12} =
M_{12}^{\rm SM} + M_{12}^{\rm SUSY}$. The mass difference is given
as $\Delta M_K = 2 |M_{12}|$, and the CP violation parameter
$|\epsilon_K| = {\rm Im} M_{12}/(\sqrt{2} \Delta M_K)$. The SM
predication for $M_{12}$ is in the fourth quadrant of the
$M_{12}$-complex plain. The experimental measurement for $M_{12}$ is
more accurate rather than the illustration in the Fig.2. However,
the numerical value can have ambiguity from bag parameters and the
charm quark mass. The experiment measures only the absolute values
of real and imaginary parts of $M_{12}$. So the possible solutions
to satisfy the experiment are the four separate regions as shown in
the Fig.2. Solution A is given as $M_{12}^{\rm SUSY} \sim - 2
M_{12}^{\rm SM}$. In this solution, $M_{12}$ is in the second
quadrant. Since ${\rm Im} M_{12} \ll {\rm Re}M_{12}$, $M_{12}$ lying
in the third quadrant is almost same as solution A. In  solution A,
the $M_{12}^{\rm SUSY}$ phase is almost $\pi$. In solution B,
$M_{12}$ is in the first quadrant, and the $M_{12}^{\rm SUSY}$ phase
is about $\pi/2$. In solution C, $M_{12}$ is in the fourth quadrant.
When $|M_{12}^{\rm SUSY}| \ll {\rm Im} M_{12}^{\rm SM}$, the SUSY
contribution is negligible in the $K$ system, and phase of
$M_{12}^{\rm SUSY}$ can be arbitrary. When $|M_{12}^{\rm SUSY}| \sim
{\rm Im} M_{12}^{\rm SM}$, the phase of $M_{12}^{\rm SUSY}$ should
be 0 or $\pi$ in  solution C. The solutions A, B, C which provide
maximal value of $\kappa$ are shown in the Fig.1. In solutions B and
C, the amount of cancellation of $\delta_{12}^d$ is larger than  in
 solution A for a given $\kappa \sim 0.2$.
%

\begin{figure}[t]
 \center
 \includegraphics[viewport = 25 24 280 220,width=8.0cm]{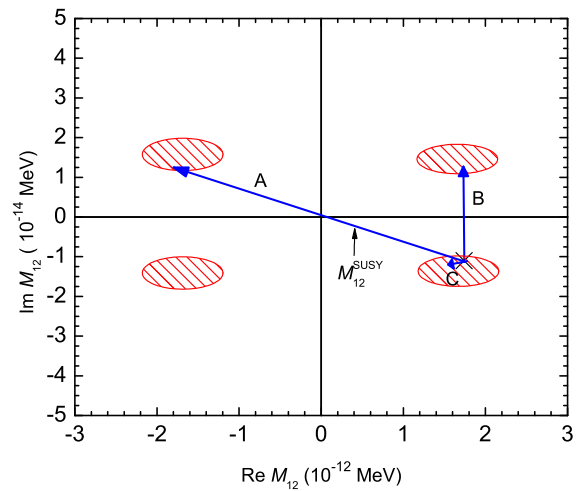}
 \caption{
Illustration of the experimentally allowed solutions (shaded area)
for amplitude $M_{12}$ for $K$-$\bar K$ mixing. Details are
described in the text. } \vspace{-0.5cm}
\end{figure}

As  noted, the phases of  SUSY contributions for $K$-$\bar K$,
$B_d$-$\bar B_d$, $B_s$-$\bar B_s$ mixing amplitudes are related
since the phases of $\delta_{ij}^d$ are cancelled (up to small RGE
modification) and only two physical phases in $P_d$ remain.
Since all solutions  A,B,C of the $K$-$\bar K$ mixing provide
restriction to the phases of the SUSY contributions, phases of the
SUSY contribution for $B_{d,s}$-$\bar B_{d,s}$ mixings are
restricted for large  SUSY contribution. We draw Fig.3
to show the correlation of the phases. We choose CKM parameters as
$\sin 2\beta^{\rm SM} \simeq 0.77$ and $\sin 2\beta_s^{\rm SM}
\simeq 0.04$. We use the same parameters for $\theta_{23}^q$,
$\theta_{12}^q$, $k_2$ and SUSY mass parameters as we have used to
draw Fig.1. We choose $\kappa = 0.2$ in each solution. As described,
the phase of the SUSY contribution $M_{12}^{K{\mbox -}\bar K}$ is
almost $\pi$ in  solution A. We choose the SUSY phases to be $\pi/2$
and $\pi$ for solutions B and C, respectively. In the plot, we
choose the absolute values of SUSY contributions to be same for both
solutions B and C.
Since the $B_s$-$\bar B_s$ SUSY contribution is determined by
$\delta_{23}^d$ (eq.(\ref{delta23})), the maximal values of
$|\beta_s|$ are almost same in all  three solutions. On the other
hand, the $B_d$-$\bar B_d$ SUSY contribution depends on
$\theta_{13}^q$ and it is different for  solutions A and B,C.
One finds that
$\sin2\beta^{\rm eff}$ is smaller than the SM value
when $\beta_s^{\rm eff}$ becomes positive
 in  solutions A and C.
Solution B gives us  opposite result. This correlation is a
consequence of the fact that there are only two physical phases for
three different mixing amplitudes.

\begin{figure}[t]
 \center
 \includegraphics[viewport = 25 24 280 220,width=8.0cm]{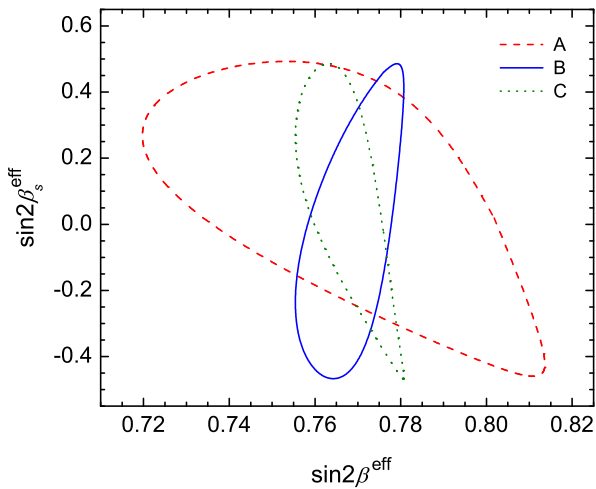}
 \caption{
Correlations for the phases of $B_d$-$\bar B_d$ ($\beta^{\rm eff})$
and $B_s$-$\bar B_s$ ($\beta_s^{\rm eff}$) mixings. A, B, C
correspond to the solutions illustrated in Fig.2. } \vspace{-0.5cm}
\end{figure}

The global fit of the experimental data
\cite{Bona:2006sa,Charles:2004jd} shows that $\sin2\beta$ arising
from the $V_{ub}$ measurement has a $2\sigma$ discrepancy from the
$\sin2\beta$ measurement from $B_d \to J/\psi K$ \cite{Bona:2006sa},
$\sin2\beta=  0.678\pm 0.026$ \cite{hfag}. Thus a negative SUSY
contribution is favored for $\sin2\beta^{\rm eff}$. The present data
for $\beta_s$, eq.(\ref{beta_s}), favors negative value. As a
result,  solution A is disfavored by the experimental result. For
solutions B and C, the phases of SUSY contributions have ambiguity.
In order to produce negative contribution for $\sin2\beta$ and
generate negative $\sin2\beta_s$, a phase of magnitude from $0$ to
$\pi/2$ is favored for the $K$-$\bar K$ SUSY contribution.

We have chosen $\theta_{23}^q = \pi/4$,
which generates the maximal SUSY contribution to $B_s$-$\bar B_s$ for a given $\kappa$.
It is important that the $D$-$\bar D$ mixing constrains
$\kappa \sin^2 \theta_{23}^q$ at the
$K$-$\bar K$ cancellation region, 
and thus, the $D$-$\bar D$ mixing data constrains the $B_s$-$\bar
B_s$ mixing for a given $\theta_{23}^q$. Naively, the SUSY
contribution of $B_s$-$\bar B_s$ is proportional to $(\kappa
\sin2\theta_{23}^q)^2$. Thus, when the SUSY contribution saturates
the observed $D$-$\bar D$ mixing, the maximal value of $|\beta_s|$
becomes larger for smaller $\theta_{23}^q$. Such a direction also
decreases $\sin2\beta^{\rm eff}$, which is favored by the
experimental result. So we see that the meson mixings are all
related in  left-right symmetric models. More accurate measurement
of $B_s$-$\bar B_s$ phase will impose interesting constraint on the
model.

If we consider a $SO(10)$ model, the SUSY contribution of
$B_s$-$\bar B_s$ also gets correlated to the $\tau \to \mu \gamma$
decay amplitude \cite{Dutta:2006gq}, which is more important
compared to the $D$-$\bar D$ constraint for small $m_0$ and large
$\tan\beta_H$. In the case of large $m_0$, however, the $D$-$\bar D$
constraint can be stronger than $\tau \to \mu \gamma$. The $\mu \to
e \gamma$ decay amplitude is small due to the same cancellation
condition for $\delta_{12}^d$(which reduces $K$-$\bar K$ mixing
amplitude) and $\delta_{12}^l$.

We assume that the left-right symmetry under the  exchange
of $Q_L \leftrightarrow Q_R^c$. We can also consider the exchange
$Q_L \leftrightarrow (Q_R^c)^*$. In this case, the Yukawa matrices
are Hermitian instead of symmetric matrices \cite{Dutta:2004hp}.
The phase matrices $P_d$ and $P_d$ become just signature matrices.
%
However, the squark mass matrices satisfy $M_U^2 = M_D^2 =
(M_Q^2)^*$ and therefore, the phases in $P$ are not cancelled in the
meson mixings.
As a result,
in the Hermitian Yukawa case,
the phases of meson mixings are also correlated
at the $K$-$\bar K$ cancellation region
as in the symmetric Yukawa case.

In conclusion, we have studied the importance of $D$-$\bar D$
mixing in left-right symmetric models.
We showed that the $D$-$\bar D$ mixing data constrains
the phase of $B$-$\bar B$ mixing for given parameters in left-right symmetric models,
and studied the correlation of the meson mixings.
If we consider unified models without left-right symmetry such as
$SU(5)$, where only right-handed squark mixings can be large
naively, the SUSY contribution is not very enhanced. Besides, since
right-handed squark mixings are unknown, both $K$-$\bar K$ and
$D$-$\bar D$ can be cancelled away separately, and therefore there
is no constraint. Therefore, left-right symmetric models are very
interesting candidates to investigate correlation among measurements
especially when the SUSY contribution is maximal and the $B_s$-$\bar
B_s$ phase is large. The improved result of this mixing phase will
further shed light on the correlation of meson mixings and
left-right models.

This work was supported in part by the DOE grant DE-FG02-95ER40917.


\end{document}